\documentclass{aa}  

\usepackage{graphicx}
\usepackage{txfonts}
\usepackage{comment}
\newcommand{\ha}          {H$\alpha$}
\newcommand{\hb}          {H$\beta$}
\newcommand{\wha}          {$W_{\rm H\alpha}$}
\newcommand{\mwhab}          {$\langle W_{\rm H\alpha, b} \rangle$}
\newcommand{\kmpers}      {\mbox{\rm km~s$^{-1}$}}
\newcommand{\kms} {\kmpers}

\newcommand{\fmol}         {\mbox{$f_{\rm mol}$}}

\newcommand{\dms}         {\mbox{$\Delta$SFMS$_{\rm g}$}}

\newcommand{\tdep}{\mbox{$\tau_{\rm dep,mol}$}}

\begin{document}

   \title{The EDGE-CALIFA survey: exploring the role of the molecular gas on the galaxy star formation quenching}

   \subtitle{}

   \author{D. Colombo\inst{1}\thanks{dcolombo@mpifr-bonn.mpg.de},
          S.F. Sanchez\inst{2},
          A. D. Bolatto\inst{3},
          V. Kalinova\inst{1},
          A. Wei\ss\inst{1},
          T. Wong\inst{4},
          E. Rosolowsky\inst{5},
          S. N. Vogel\inst{3},\\          
          J. Barrera-Ballesteros\inst{2},
          H. Dannerbauer\inst{6,7},
          Y. Cao\inst{8},
          R. C. Levy\inst{3},
          D. Utomo\inst{9},
          L. Blitz\inst{10}
          }

   \institute{Max-Planck-Institut f\"ur Radioastronomie, Auf dem H\"ugel 69, 53121 Bonn, Germany
   \and
   Instituto de Astronomi\'a, Universidad Nacional Auton\'oma de Mexico, A.P. 70-264, 04510 M\'exico, D.F., Mexico
   \and
   Department of Astronomy, University of Maryland, College Park, MD 20742, USA
   \and
   Department of Astronomy, University of Illinois, Urbana, IL 61801, USA
   \and
   Department of Physics, University of Alberta, 4-181 CCIS, Edmonton, AB T6G 2E1, Canada
   \and
   Instituto de Astrof\'isica de Canarias, E-38205 La Laguna, Tenerife, Spain
   \and
   Universidad de La Laguna, Dpto. Astrof\'isica, E-38206 La Laguna, Tenerife, Spain
   \and
   Aix Marseille Univ, CNRS, CNES, LAM (Laboratoire d’Astrophysique de Marseille), Marseille,France
   \and
   Department of Astronomy, The Ohio State University, 140 West 18$^{\rm th}$ Avenue, Columbus, OH 43210, USA
   \and
   Department of Astronomy, University of California, Berkeley, CA 94720, USA
   }

   \date{Received XXX; accepted XXX}

 
  \abstract
    {Understanding how galaxies cease to form stars represents an outstanding challenge for galaxy evolution theories. This process of ``star formation quenching'' has been related to various causes, including Active Galactic Nuclei (AGN) activity, the influence of large-scale dynamics, and the environment in which galaxies live. In this paper, we present the first results from a follow-up of ``Calar Alto Legacy Integral Field Area'' (CALIFA) survey galaxies with observations of molecular gas obtained with the ``Atacama Pathfinder Experiment'' (APEX) telescope. Together with the ``Extragalactic Database for Galaxy Evolution'' (EDGE) survey ``Combined Array for Research in Millimeter-wave Astronomy'' (CARMA) observations, we collect $^{12}$CO observations that cover approximately one effective radius in 472 CALIFA galaxies. We observe that the deficit of galaxy star formation with respect to the star formation main sequence (SFMS) increases with the absence of molecular gas and with a reduced efficiency of conversion of molecular gas into stars, in line with results of other integrated studies. However, by dividing the sample into galaxies dominated by star formation and galaxies quenched in their centres (as indicated by the average value of the \ha\ equivalent width), we find that this deficit increases sharply once a certain level of gas consumption is reached, indicating that different mechanisms drive separation from the SFMS in star-forming and quenched galaxies. 
    Our results indicate that differences in the amount of molecular gas at a fixed stellar mass are the primary driver for the dispersion in the SFMS, and the most likely explanation for the start of star-formation quenching. However, once a galaxy is quenched, changes in star formation efficiency drive how much a retired galaxy separates in star formation rate from star-forming ones of similar masses. In other words, once a paucity of molecular gas has significantly reduced star formation, changes in the star formation efficiency are what drives a galaxy deeper into the red cloud, retiring it.}

   \keywords{ISM: molecules --
                Galaxies: evolution --
                Galaxies: ISM --
                Galaxies: star formation
               }

   \titlerunning{Molecular gas and star formation quenching in CALIFA galaxies}

   \authorrunning{D. Colombo, S. Sanchez, A. Bolatto et al.}

   \maketitle

\section{Introduction}\label{S:introduction}
The appearance of galaxies in the nearby Universe is largely shaped by their star formation activity. The cessation of star formation that accompanies the transformation of a blue, spiral galaxy into a ``red-and-dead'' elliptical is usually called ``star formation quenching'' \citep[e.g., ][]{faber2007}. Quenching is generally associated with the shortage of the raw fuel that feeds the star formation: the cold gas, and in particular its molecular phase. In low-mass galaxies, the gas, which is weakly bound due to their shallow potential wells, can be promptly removed by stellar feedback \citep[e.g., ][]{dekel_silk1986}. High-mass galaxies, instead, might require a more powerful way to disperse the gas, such as Active Galactic Nuclei (AGN) outflows \citep[e.g., ][]{lacerda2020}. The AGN activity, by heating up the gas, can also block the accretion from the intergalactic medium causing ``quenching by starvation'' \citep[e.g., ][]{cicone2014}. Likewise, the suppression of cold gas accretion can result from shock-heating in dark matter halos with mass $>10^{12}$\,M$_{\odot}$ (``halo quenching''; e.g., \citealt{dekel_birnboim2006}). Small galaxies falling towards a galaxy cluster can have their gas removed by tidal stripping \citep[e.g., ][]{abadi1999}, or through the interaction with hot intra-cluster medium \citep[e.g., ][]{moore1996} that prevents further accretion from the intergalactic medium (``environmental quenching''). Alternatively, galaxies can stop forming stars efficiently even if a substantial amount of gas is present. In this ``morphological quenching'' scenario the development of a bulge or a spheroid, together with dispersive forces such as shear, stabilises the galactic gaseous disk against collapse, preventing star formation \citep{martig2009}. To fully disentangle the dominant quenching mechanisms, their time-scales, and their parameter dependencies requires the analysis of cold gas conditions for a statistically significant sample of galaxies.

Major nearby galaxy cold gas mapping surveys (\citealt{regan01}, \citealt{wilson2009}, \citealt{rahman2011}, \citealt{leroy09}, \citealt{donovan_meyer2013}, \citealt{bolatto2017}, \citealt{sorai2019}, \citealt{sun2018}) 
have focused on observations of the molecular gas (through CO lines). Despite a few notable exceptions \citep[e.g., ][]{alatalo2013,saintonge2017}, these surveys observed mainly spiral or infrared-bright galaxies (i.e.  galaxies with significant star formation) and have emphasized understanding of how star formation happens, rather than how it stops. This boils down to quantifying the relation between molecular gas and star formation rate (SFR), which appears nearly linear in nearby disks \citep{kennicutt1998,bigiel08,leroy2013,lin2019}. This relationship is often parametrized via the ratio between the SFR and the molecular gas mass, called ``molecular star formation efficiency'' (SFE=SFR/$M_{\rm mol}=1/\tau_\mathrm{dep}$), where the inverse of the SFE is the ``depletion time,'' $\tau_{\rm dep}$.  The depletion time indicates how much time is necessary to convert all the available molecular gas into stars at the current star formation rate. On kpc scales and in the disks of nearby, star-forming galaxies, $\tau_{\rm dep}$ is approximately constant around 1-2\,Gyr \citep{bigiel2011, rahman2012, leroy2013, utomo2017}, and it appears to weakly correlate with many galactic properties such as stellar mass surface density or environmental hydrostatic pressure \citep{leroy2008, rahman2012}. Nevertheless, small but important deviations for a constant SFE have been noticed, which can be the first hints of star formation quenching. In some galaxies, the depletion time in the centres appear shorter \citep{leroy2013, utomo2017} or longer \citep{utomo2017} with respect to their disks. These differences may correlate with the presence of a bar or with galaxy mergers  (\citealt{utomo2017}; see also \citealt{muraoka2019}) and do not seem to be related to unaccounted variation in the CO-to-H$_2$ conversion factor \citep{leroy2013,utomo2017}. Spiral arm streaming motions have also been observed to lengthen depletion times \citep{meidt13, leroy2015}. 

Besides variation of the SFE within galaxies, differences in global SFE between galaxies have been explored more widely in the nearby Universe, thanks especially to molecular gas galaxy-integrated studies \citep[see ][and references therein]{saintonge2011,saintonge2017}. Those studies are less expensive in terms of exposure time compared to resolved mapping studies and provide the opportunity to collect data for larger galactic samples. In particular, integrated samples provide access to the molecular gas content of galaxies below the ``star formation main sequence'' \citep[SFMS or MS, i.e the locus of the star-forming galaxies in the SFR-stellar mass diagram e.g., ][]{brinchmann2004,whitaker2012,renzini_peng2015, cano_diaz2016}, that is, galaxies that are slowly shutting down their star formation (located in the so-called ``green valley''; \citealt{salim2007}), down to passive galaxies (or ``red sequence'' galaxies, as defined on a colour-magnitude diagram). In general, galaxies on the main sequence have longer depletion times than similar stellar mass galaxies located below the main sequence \citep{saintonge2016, saintonge2017}. The reasons for this are unclear and might be due to a combination of effects. Barred and interacting galaxies show generally shorter global depletion times compared to other systems \citep{saintonge2012}. Early-type galaxies show lower SFE compared to late-type objects \citep{davis2014} and some of the lowest values of SFE are observed in bulge-dominated galaxies \citep{saintonge2012}. As star formation quenching is more often seen to happen inside-out \citep[e.g., ][]{gonzalez_delgado2016}, galaxy morphology and structural properties seem to play a role in modifying the SFE. On average, \tdep\ appears to decrease moving from early- to late-type systems \citep{colombo2018}, following the decrease in shear \citep{davis2014, colombo2018} as described by the ``morphological quenching'' scenario \citep[but see ][]{koyama2019}. The molecular depletion time is also seen to decrease with stellar surface density and increase with molecular gas velocity dispersion \citep{dey2019}. This might indicate that \tdep\ is longer for bulged systems, and where gas is less gravitationally bound (as the gas boundness is proportional to $\Sigma_{\rm mol}/\sigma_{\rm CO}$, i.e. the ratio between the molecular gas mass surface density and the CO velocity dispersion; see \citealt{leroy2015}). The environment in which a galaxy lives might also be important: galaxies in clusters appear to have longer depletion time than group galaxies \citep[e.g., ][]{mok2016}, possibly due to the turbulent pressure and additional heating induced by the cluster itself. 

Nevertheless, most theories attribute star formation quenching to the absence of molecular gas, rather than to a less efficient conversion from gas to stars. This has been explored observationally mostly by integrated surveys \citep{genzel2015,saintonge2016,tacconi2018,lin2017}, which usually parameterised the shortage of gas through the molecular gas fraction, $f_{\rm mol}=M_{\rm mol}/M_*$. This quantity seems to drop drastically for galaxies with $M_*>10^{10.5-11}$\,M$_{\odot}$ \citep{saintonge2017,bolatto2017} and for redder objects \citep{saintonge2011}. It appears also reduced in barred galaxies \citep{bolatto2017}. This absence appears tentatively connected to the presence of an AGN in a galaxy \citep{saintonge2017}; negative feedback due to the AGN may provide an efficient gas-removal mechanism, which is suggested by the fact that AGN-hosting galaxies are almost exclusively observed in the green valley (see \citealt{lacerda2020}). Nevertheless, this is still matter of debate \citep[e.g., ][]{kirkpatrick2014,rosario2018}. 

Despite several studies exploring the methods for quenching, there is not a clear conclusion as to whether the quenching is driven by the reduction in molecular gas content, a change in the star formation efficiency of the molecular gas or both effects. Furthermore, these studies have not assessed how these effects may change throughout the quenching process. To address these shortcomings, in this work, rather than examining causes for SFE and \fmol\ variations, we use the $^{12}$CO(1-0) maps from EDGE \citep{bolatto2017} in combination with new $^{12}$CO(2-1) single-dish measurements to investigate whether SFE or $f_{\rm mol}$ changes are the main cause of star formation quenching in the centre of more than 470 CALIFA (\citealt{sanchez2016}) galaxies at different quenching stages. The paper is structured as follows. Section~\ref{S:data} exposes the data used in this paper, while Section~\ref{S:props} describes the quantities derived from these data such as star formation rates (SFRs) and molecular gas masses ($M_{\rm mol}$). The results of the analyses are shown in Section~\ref{S:results}. Those results are discussed and summarised in Section~\ref{S:summary}. For the derived quantities we assume here a cosmology $H_0 = 71\,$km$\,$s$^{-1}\,$Mpc$^{-1}$, $\Omega_{\rm m}$ =0.27,  $\Omega_{\rm \Lambda}$ =0.73.

\begin{figure}
    \centering
    \includegraphics[width = 0.4 \paperwidth, keepaspectratio]{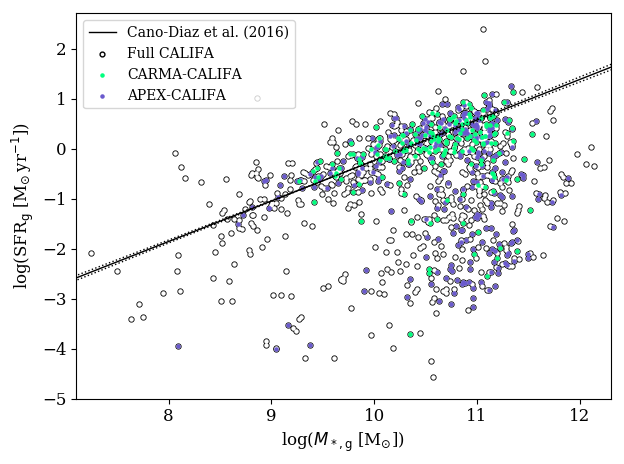}
    \caption{Star formation rate vs stellar mass integrated over each galaxy, comparing the distributions of galaxies from the CARMA and APEX subsets and the remaining CALIFA galaxies. The star formation main sequence is indicated using the Cano-D\`{i}az et al. (2016) fit (full black line) with its confidence level (dotted black lines). The diagram is zoomed-in to emphasise the CARMA and APEX coverage. The extended CALIFA sample has SFR=$10^{-6.2}-10^{3.3}$\,M$_{\odot}$\,yr$^{-1}$ and $M_*=10^{5.7}-10^{13.7}$\,M$_{\odot}$, however only a few objects have star formation rates and stellar mass outside the range shown in the figure.}
    \label{F:samples}
\end{figure}

\section{Sample and data}\label{S:data}
We collect a homogenised compilation of 472 galaxies with molecular gas measured in an aperture of diameter 26.3" (corresponding to the APEX beam at 230\,GHz). This compilation comprises our new observations using APEX together with a re-analysis of CARMA observations acquired by the EDGE collaboration \citep{bolatto2017}. All galaxies were already covered by spatially resolved IFS observations by the CALIFA survey \citep{sanchez2012}. The sample of 472 galaxies covers a wide range of galaxy parameters in terms of morphology (from E to Sm, including a few irregular galaxies), stellar masses ($10^{7.2}-10^{11.9}$\,M$_{\odot}$), and star-formation rates ($10^{-4.0}-10^{1.3}$\,M$_{\odot}$\,yr$^{-1}$). Thus, it is one of the first explorations of a large sample not systematically biased by selection. The CARMA sample is generally made up of galaxies concentrated along the SFMS (as it has been assembled considering 22$\mu$m bright WISE galaxies), while the APEX sample targets cover more uniformly the so-called ``green valley'' and ``red sequence'', as we can see in Fig.~\ref{F:samples}. Together, the CARMA and APEX samples provide good coverage of the full extended CALIFA sample. The APEX and CARMA samples do not overlap, meaning that objects observed by CARMA in $^{12}$CO(1-0) have not been re-observed with APEX in $^{12}$CO(2-1).

\begin{figure*}[h!]
    \centering
    \includegraphics[width = 0.85 \paperwidth, keepaspectratio]{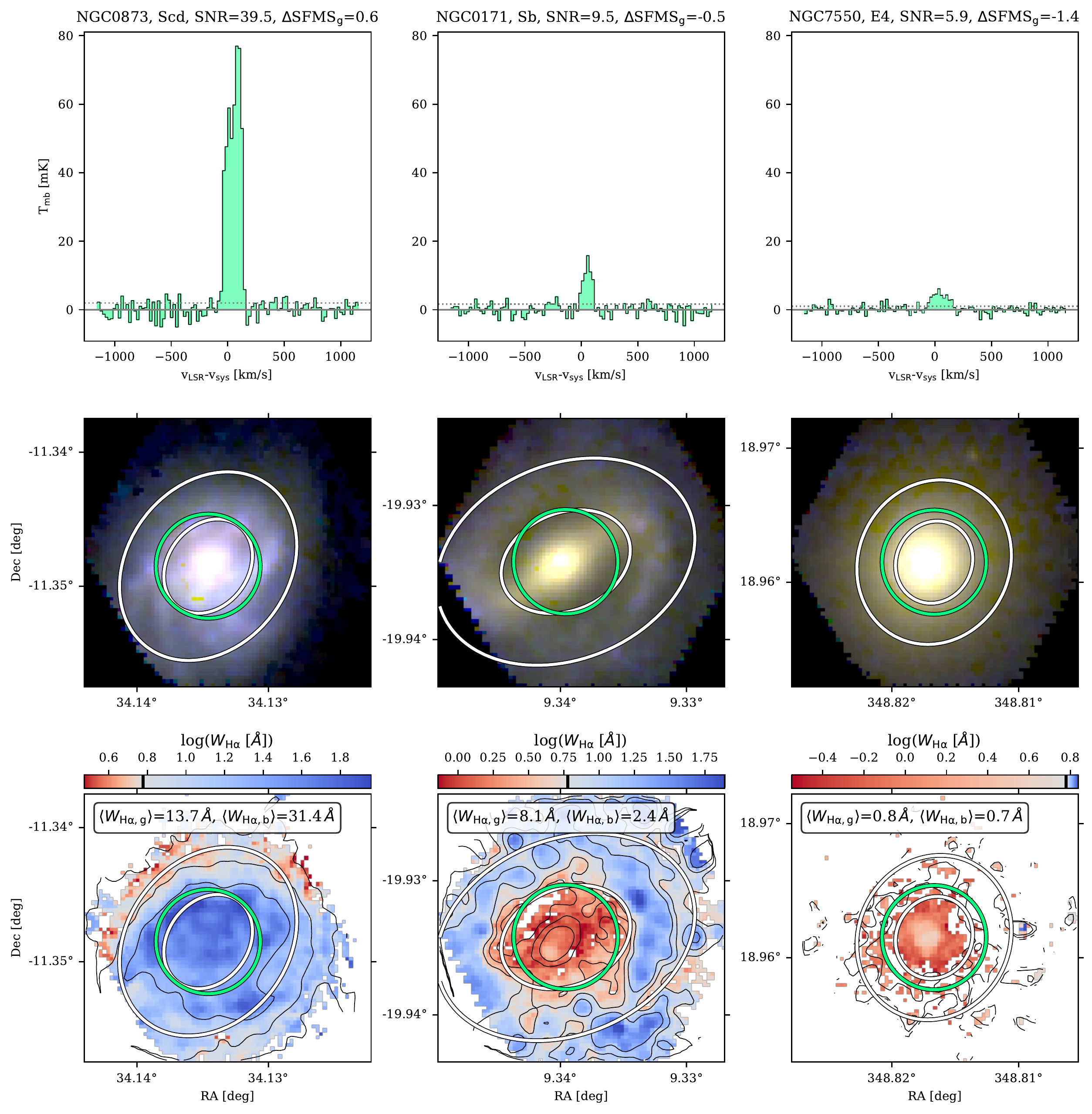}
    \caption{APEX $^{12}$CO(2-1) spectra of three observed galaxies in different quenching phases. Top row panels show the spectra for each galaxy in green, where the dotted line represents the observation $\sigma_{\rm RMS}$. In the \emph{top row} panel titles, the name of the galaxy (as in the CALIFA database) and its morphology, CO signal-to-noise ratio (SNR), and the logarithmic ratio of the galaxy global star formation rate to the global star formation main sequence ($\Delta$SFMS$_{\rm g}$) are shown. \emph{Middle row} panels show continuum RGB images extracted from the CALIFA datacubes using $u-$ (blue), $g-$ (green) and $r-$ (red) bands. In the \emph{bottom row}, the \wha\ maps are displayed in diverging red colours where $W_{\rm H\alpha}<6\AA$, while the diverging blues illustrate the part of the map where $W_{\rm H\alpha}>6\AA$. The colour maps are centred at $W_{\rm H\alpha}=6\AA$ in logarithmic units, and this value is indicated as a black vertical line in the colour-bars. Black contours mark the 25$^{\th}$, 50$^{\th}$, and 75$^{\th}$ percentiles of the $\log(F_{\rm H\alpha})$ distribution, previously masked at 3$\sigma_{\rm RMS}$. \wha\ map is also masked below 3$\sigma_{\rm RMS}$ of \ha\ flux map. In panel legends, the median \ha\ equivalent width across the whole map ($\langle W_{\rm H\alpha, g}\rangle$) and within the beam aperture ($\langle W_{\rm H\alpha,b}\rangle$) are presented. In the panels of the two bottom rows, the green circle shows the APEX beam (FWHM=26.2 arcsec at 230\,GHz), while white ellipsoids indicate 1 and 2\,$R_{\rm eff}$.}
    \label{F:mapspectra}
\end{figure*}

Fig.~\ref{F:mapspectra} gives an example of the quality of the APEX data and the rich variety of the targets in our dataset. Spectra of APEX $^{12}$CO(2-1) observations are illustrated in the first row. The centres of all the galaxies in this example are well detected in CO, but the continuum and \ha\ equivalent width (\wha) maps show that the objects are in three different phases of their evolution. On the left middle panel, the continuum map indicates that the galaxy (NGC0873) is entirely blue, therefore dominated by star formation. Along with colour,  \wha\ is also a faithful proxy for the star formation properties of the galaxies \citep[e.g., ][]{sanchez2020}. In particular, $W_{H\alpha}>6\AA$ is found in galaxy areas dominated by HII regions, while where $W_{H\alpha}<6\AA$ the galaxy is quenched, or dominated by other effects than recent star formation. Indeed, the bottom left panel of Fig~\ref{F:mapspectra} indicates that $W_{H\alpha}>6\AA$ almost everywhere and closely follows the continuum map information. In NGC0170 (middle column), instead, the star formation in the centre is fully quenched, as indicated by the median value of \wha\ within the APEX beam aperture, $\langle W_{\rm H\alpha, b}\rangle <6\AA$ (see Section~\ref{S:props} for further details), and by the yellow colour of the continuum map in the central region. Nevertheless, on the outskirts, stars are forming, and this galaxy appears globally dominated by star formation (as suggested by the median \wha\ across the full galaxy $\langle W_{\rm H\alpha, g} \rangle >6\AA$). The last galaxy displayed in the figure, NGC7550, is fully retired, as suggested by  $W_{\rm H\alpha}<6\AA$ basically everywhere and by the yellow colour of the whole continuum map. The galaxy, however, still possesses a measurable amount of molecular gas. In the following we assume as ``centrally star-forming'' galaxies the objects that show a median $W_{\rm H\alpha}>6\AA$ within the APEX beam, and ``centrally retired, quenched or quiescent'' the targets where median $W_{\rm H\alpha}<6\AA$ within the APEX beam.

\subsection{CALIFA data}\label{SS:califa}
CALIFA is an integral field spectroscopy (IFS) optical survey that imaged more than 1000 galaxies (667 included in the data release 3, and 416 in the extended sample) using the PMAS/PPak integral field unit instrument mounted on the 3.5m telescope of the Calar Alto Observatory \citep{sanchez2012,sanchez2016,lacerda2020}.
The CALIFA sample is drawn from the Sloan Digital Sky Survey (SDSS, \citealt{york2000}) to reflect the present-day galaxy population (0.005$<z<$0.03) in a statistically meaningful manner (log($M_*/$[M$_{\odot}$])=9.4-11.4; E to Sd morphologies, including irregulars, interacting, and mergers; \citealt{walcher2014,barrera-ballesteros2015}). Here we consider the galaxies observed with the low-resolution (V500) setup which covers between 3745–7500$\,\AA$ with a spectral resolution  FWHM=6$\,\AA$.  CALIFA datacubes possess a spatial resolution of FWHM$\sim2.5$ arcsec \citep{garcia-benito2015}. Given the limits on the redshift, CALIFA allows the study of galaxies on kpc-scale. Additionally, the maps extend beyond 2.5\,$R_{\rm eff}$, covering most of the optical disks. Ionized gas and stellar continuum map properties have been obtained through the {\sc Pipe3D} pipeline \citep{sanchez2016b, sanchez2016c}. {\sc Pipe3D} analyses the stellar population applying the GSD156 simple stellar populations (SSP) library \citep{cid_fernandes2013}. A stellar population fit is performed to the spatially rebinned V-band datacubes in order to estimate a spaxel-wise stellar population model. This model is used to calculate the stellar mass density value within each spaxel. The ionised gas datacube is then generated by subtracting the stellar population model from the original cube. Each of the 52 sets of emission line maps is performed, calculating flux intensity, centroid velocity, velocity dispersion, and equivalent width for every single spectrum.

\subsection{APEX observations and survey goal}\label{SS:observations}
We observed the $^{12}$CO(2-1) emission (rest frequency, $\nu_{\rm ^{12}CO(2\textendash1)}$ = 230.538\,GHz) from 296 galaxy centres and 39 off-centre positions. In this paper, we present only the centre observations. The project was carried out with the APEX 12\,m sub-millimetre telescope \citep{guesten2006} in ON-OFF mode using the wobbler (which ensures stable baselines), and the PI230 receiver which operates in the 1.3 mm atmospheric window. Galaxies have been observed across two projects M9518A\_130 and M9504A\_104 (PI: D. Colombo) which allocated  180 and 205 hours in the summer and winter semesters 2019, respectively, for a total of approximately 385 hours which include calibrations, additional overheads, and further test observations. All galaxies have been drawn from the CALIFA extended sample, with the only requirement to be accessible by APEX, i.e. all galaxies in the sample have declination $\leq30^{\circ}$.

The APEX resolution at 230 GHz is 26.3 arcsec. The median ratio of the beam radius to the effective radius of the full sample of galaxies is 1.12, with an inter-quartile range of 0.60. These do not change much if we consider only the face-on targets (with an inclination less than 65$^{\circ}$). This means that, on average, the APEX beam covers roughly half of the radial extent of the CALIFA maps (see Section~\ref{SS:califa} and Fig.~\ref{F:mapspectra}).

The survey is designed to reach a uniform rms of 2\,mK (70\,mJy) per $\delta v=30$\,\kms\ wide-channels. For several targets for which we have detections but low signal-to-noise ratio (SNR<3), we integrate longer to achieve a rms of 1\,mK. These requirements allowed us to attain pointed observations 10$\times$ more sensitive than CARMA (in term of achievable minimum molecular gas mass surface density; see also Section~\ref{SS:carma}) and thereby detect the CO line in even the most gas-poor galaxies, which constituted the main goal of our APEX observations. In particular, we obtain 207 CO detections (SNR$\ge3$) for a detection rate of 70\%, with 50\% of the observed galaxies detected at $5\sigma_{\rm RMS}$. 

Data calibration and reduction of the APEX data have been performed using the ``Grenoble Image and Line Data Analysis Software'' (GILDAS\footnote{http://www.iram.fr/IRAMFR/GILDAS}) and ``Continuum and Line Analysis Single-dish Software'' (CLASS) package with which we fit and remove a linear baseline to each spectrum outside a window of 600\,\kms\, centred on the the galaxy V$_{\rm LSR}$. Afterwards, we smooth the data to a common spectral resolution $\delta v=23$\,km\,s$^{-1}$. The final median rms from the full sample at 23\,km\,s$^{-1}$ is 2.2 mK, which corresponds to a median, 3$\sigma_{\rm RMS}$, $M_{\rm mol}=2.8\times10^8$\,M$_{\odot}$ ($\Sigma_{\rm mol}\sim2.8$\,M$_{\odot}$\,pc$^{-2}$) at the median distance of the sample of $\sim67\,$Mpc and using a constant $\alpha_{\rm CO(2-1)}=6.23$\,M$_{\odot}$\,(K\,km\,s$^{-1}$\,pc$^2$)$^{-1}$. Full details about survey specifics and data reduction will be presented in an upcoming survey paper.

\subsection{CARMA data}\label{SS:carma}
In addition to the APEX data, in this paper, we also make use of the CARMA database which constitute the first $^{12}$CO(1-0) (and $^{13}$CO(1-0)) follow-up of CALIFA galaxies undertaken by the EDGE collaboration. A full description of the CARMA CO data is given in \cite{bolatto2017}; here we provide a brief summary. The EDGE-CALIFA collaboration originally mapped 177 infrared-bright CALIFA galaxies with CARMA E-configuration. A sub-sample of 126 higher signal-to-noise galaxies were observed also in D-configuration; subsequently, for these galaxies, D+E combined cubes were produced. In this paper, we use the $^{12}$CO(1-0) data of the 126 D+E galaxies and the remaining 51 E-configuration galaxies that were not followed-up with the D-array of CARMA. The data cubes were smoothed to a spectral resolution of 20\,km\,s$^{-1}$. The final D+E galaxies have 4.5 arcsec resolution, while the E-configuration only galaxies show a 9 arcsec resolution. The average rms noise in the D+E galaxies is 38 mK per 20\,km\,s$^{-1}$ channel width. This dataset was extensively exploited in \cite{utomo2017, colombo2018, levy2018, leung2018, chown2019, levy2019, dey2019, barrera-ballesteros2020}, and recently reviewed in \cite{sanchez2020}, exploring different aspects of the interconnection between the ionised and cold molecular gas phases in galaxies.

\section{Derived quantities}\label{S:props}
In this paper, we use spatially unresolved (i.e. single beam) data from APEX as well as resolved data from CARMA and CALIFA. In order to make these data comparable and simulate the effect the APEX beam would have on CARMA and CALIFA maps, we introduce a ``tapering'' function $W_{\rm T}$, i.e. a bi-dimensional Gaussian, centred on the centre of the galaxy, with unitary amplitude and FWHM $\theta$= $\sqrt{\theta_{\rm APEX}^2-\theta_{\rm CARMA, CALIFA}^2}$, where the APEX beam FWHM $\theta_{\rm APEX}=26.3\,$arcsec, while the CARMA beam FWHM $\theta_{\rm CARMA}=4.5\,$arcsec or $\theta_{\rm CARMA}=9\,$arcsec, for the D+E or E-only configuration, respectively; and $\theta_{\rm CALIFA}=2.5\,$arcsec. Integrated quantities within the APEX beam aperture will be indicated with the sub-script ``b'' and are calculated by co-adding the pixels within the CARMA or CALIFA maps, previously multiplied by the Gaussian filter, $W_{\rm T}$. Average quantities within the APEX beam aperture are obtained using the weighted median of pixels values in the resolved maps where the weights for each pixel are given by the ``tapering'' function, $W_{\rm T}$.  This operation is equivalent to the convolution of the CALIFA property maps to the APEX beam size and sampling the result at the pointing centre of the APEX beam. Globally integrated quantities, denoted with the subscript ``g'', are measured by summing up all the pixels in a given map, without applying the Gaussian taper. Where no subscript is indicated, the quantity is computed spaxel or pixel-wise for CALIFA and CARMA maps, respectively.

\subsection{CO luminosity from ON-OFF APEX observations}\label{SS:props_co_apex}

We derive the $^{12}$CO(2-1) flux within a spectral window of 400\,km\,s$^{-1}$ centred on the systemic velocity of the galaxy which is derived from the stellar redshift. The CO line velocity-integrated flux is expressed by the following equation:

\begin{equation}\label{E:fco}
    S_{\rm CO, b}\,\mathrm{[K\,km\,s^{-1}]} = \sum_i T_{{\rm mb},i}\delta v 
\end{equation}

where $\delta v=23\,$\kms\ is the data final channel width, using $T_{\rm mb}$ = $T_{\rm A}^*/\eta_{\rm mb}$, with $\eta_{\rm mb}=0.78$ for the APEX beam efficiency at 230\,GHz. $S_{\rm CO, b}$ is converted to Jy\,km\,s$^{-1}$ through  a conversion factor between Kelvin and Jansky\footnote{http://www.apex-telescope.org/telescope/efficiency/} of Jy/K=37. 

The statistical error for the flux is given by:

\begin{equation}\label{E:eps_co}
    \epsilon_{\rm CO, b}\,\mathrm{[Jy\,km\,s^{-1}]} = \sigma_{\rm RMS}\sqrt{W_{50} \delta v}. 
\end{equation}

where $\sigma_{\rm RMS}$ is the standard deviation of the flux variations measured in the first and last 20 line-free channels of each spectrum: i.e., measured in two spectral windows of 20 channels before and after the velocity range used to measured emission line intensities. Finally, $W_{50}$ is the full width at half maximum derived as $W_{50} = \sqrt{8\log(2)}\sigma_{\rm v}$, where $\sigma_{\rm v}$ is the second moment calculated in the spectral window selected to measure the emission line (i.e,. the 400 \kms\ range centred on the systemic velocity of a galaxy). The values of $W_{50}$ obtained in this way coincides well with the ones derived using the ``two slopes method'' presented by \cite{springob2005} and used elsewhere \citep{saintonge2011,saintonge2017,cicone2017}. For non-detected galaxies, or galaxies showing SNR$<3$ we make use of $\epsilon_{\rm CO, b}$ to provide an upper limit for flux given by $3\epsilon_{\rm CO, b}$, where we assume a constant $W_{50}=200$\,km\,s$^{-1}$. This value was derived using the procedure described before for $W_{50}$ resulting from stacking the central APEX spectra of all the APEX galaxies. This CO flux upper limit serves as a basis for calculating all other CO properties of non-detected galaxies presented here.

From the CO flux, we derived the $^{12}$CO(2-1) luminosity using equation 3 of \cite{solomon1997}:

\begin{equation}\label{E:lco21}
    L_{\rm CO, b}\,\mathrm{[K\,km\,s^{-1}\,pc^2]} = 3.25\times10^7\frac{D_L^2}{\nu_{\rm obs}^2 (1+z)^3}S_{\rm CO, b}
\end{equation}

where $D_L$ is the luminosity distance in Mpc (derived from the stellar redshift, $z$), $\nu_{\rm obs}$ is the observed frequency of the emission line in the rest frame in GHz, and $S_{\rm CO, b}$ is the CO velocity-integrated flux derived using equation~\ref{E:fco} (but in Jy\,km\,s$^{-1}$).

\subsection{CO luminosity from CARMA data cubes}\label{SS:props_co_carma}

The $^{12}$CO(1-0) CARMA observations in the original EDGE database comes as position-position-velocity data cubes. Therefore, we use the tapering function, $W_{\rm T}$, here. For the CARMA data, the CO velocity-integrated intensity within the APEX beam aperture is given by:

\begin{equation}\label{E:ico10}
    I_{\rm CO, b}\,\mathrm{[K\,km\,s^{-1}]} = \sum_i I_{{\rm CO},i} \times W_{\rm T}(x_i, y_i),
\end{equation}
where the summation runs on the bi-dimensional pixels of the integrated intensity map $I_{\rm CO}$ of the whole galaxy.

After this, the CO luminosity within the APEX beam aperture is calculated by:
\begin{equation}\label{E:lco10}
    L_{\rm CO, b}\,\mathrm{[K\,km\,s^{-1}\,pc^2]} = I_{\rm CO, b}\,\delta x\, \delta y\, (1+z)^{-1},
\end{equation}
where $\delta x$, $\delta y$ are the pixel sizes in pc, and $z$ is the galaxy redshift.

As in Section~\ref{SS:props_co_apex}, we provide an upper limit for the CO flux-related quantities of the non-detected galaxies (SNR$<3$) as 3$\epsilon_{\rm CO}$ given by equation~\ref{E:eps_co}, where we assume a constant $W_{50}$=200\,\kms\ . For detected targets, the full width at half maximum, $W_{50}$, of the line that enters in the calculation of the flux statistical error, $\epsilon_{\rm CO}$, is obtained for the CARMA galaxies from the CO spectrum built using by the integrated flux in each channel map.

\subsection{Molecular gas mass and $\alpha_{\rm CO}$}

The molecular gas mass follows from the CO luminosity by assuming a CO-to-H$_2$ conversion factor, $\alpha_{\rm CO, b}$:

\begin{equation}
    M_{\rm mol, b} = \alpha_{\rm CO, b}L_{\rm CO, b}
\end{equation}

For the CARMA $\rm CO(1-0)$ data we use $\alpha_{\rm CO, b}\equiv\alpha_{\rm CO(1-0), b}$ \citep{bolatto2017}. However, for the APEX observations,  an additional correction factor is required, with $\alpha_{\rm CO, b}=\alpha_{\rm CO(1-0), b}/R_{21}$, where $R_{21}$ is the CO(2-1)/CO(1-0) ratio. This value is determined to be $\sim0.7$ in nearby galaxies \citep{leroy2013,saintonge2017}.

Given that our sample consists of a variety of galaxies often quite far  from the star formation main sequence, here we assume a variable $\alpha_{\rm CO(1-0)}$ based on \cite{bolatto2013}, equation 31:

\begin{equation}\label{E:aco}
    \alpha_{\rm CO(1-0)}\,\mathrm{[M_{\odot}\,(K\,km\,s^{-1}\,pc^{2})^{-1}]} = 2.9\exp\left(\frac{0.4}{Z'}\right) \left(\frac{\Sigma_{\rm *}}{100\,\mathrm{M_{\odot}\,pc^{-2}}}\right)^{-\gamma} 
\end{equation}

\noindent where $Z'$ is the gas-phase metallicity relative to solar metallicity and $\Sigma_{\rm *}$ is the stellar mass surface density measured at each pixel in the CALIFA data and $\gamma=0.5$ where $\Sigma_{\rm *}>100$\,M$_{\odot}$\,pc$^{-2}$ or $\gamma=0$ otherwise. Unlike \cite{bolatto2013}, to avoid iterative solving here we simply assume that $\Sigma_{\rm total}\equiv\Sigma_{\rm *}$, since for our sample galaxies the gas mass surface density is generally one order of magnitude lower than the stellar mass surface density. Also, $\Sigma_{\rm GMC}^{100}$, the Giant Molecular Cloud (GMC) molecular gas mass surface density in units of 100 M$_{\odot}$\,pc$^{-2}$,  does not appear in equation~\ref{E:aco} as we assume $\Sigma_{\rm GMC}^{100} = 1$ here, considering that GMC molecular gas mass surface density inner regions of nearby galaxies and Milky Way is largely consistent with 100\,M$_{\odot}$\,pc$^{-2}$ \citep[see ][]{sun2018,colombo2019}. For galaxies where optical emission lines remain undetected we assume, we assume $Z'=1$. Details on the $Z'$ and $\Sigma_{\rm *}$ calculation are given in Section~\ref{SS:props_optical}. In equation \ref{E:aco}, $\alpha_{\rm CO}$ is calculated across the whole CALIFA map; within the APEX beam aperture, we used the weighted median from the CO-to-H$_2$ conversion factor map, $\alpha_{\rm CO, b}$, where the weights for each pixel are given by the ``tapering'' function, $W_{\rm T}$. 

Using this method we derive a median $\mu_{\alpha_{\rm CO(2-1), b}}=3.93$\,M$_{\odot}$\,(K\,km\,s$^{-1}$\,pc$^2$)$^{-1}$ with an inter-quartile range $\sigma_{\alpha_{\rm CO(2-1), b}}=2.04$\,M$_{\odot}$\,(K\,km\,s$^{-1}$\,pc$^2$)$^{-1}$ for the new dataset observed with APEX. In contrast, for the CARMA dataset we obtain $\mu_{\alpha_{\rm CO(1-0), b}}=2.76$\,M$_{\odot}$\,(K\,km\,s$^{-1}$\,pc$^2$)$^{-1}$ and $\sigma_{\alpha_{\rm CO(1-0), b}}=1.23$\,M$_{\odot}$\,(K\,km\,s$^{-1}$\,pc$^2$)$^{-1}$. Those values are few times lower than the canonical $\alpha_{\rm CO(1-0)}=4.35$\,M$_{\odot}$\,(K\,km\,s$^{-1}$\,pc$^2$)$^{-1}$ and $\alpha_{\rm CO(2-1)}=6.21$\,M$_{\odot}$\,(K\,km\,s$^{-1}$\,pc$^2$)$^{-1}$ of the Milky Way, possibly due to the fact that the stellar mass surface density in the centre of our sample galaxies (which extend to massive red sequence galaxies) is typically higher than in the Milky Way or generally star-forming galaxies, while the gas-phase metallicity of most of our galaxies is close to solar.

\begin{figure*}[h!]
    \centering
    \includegraphics[width = 0.85 \paperwidth, keepaspectratio]{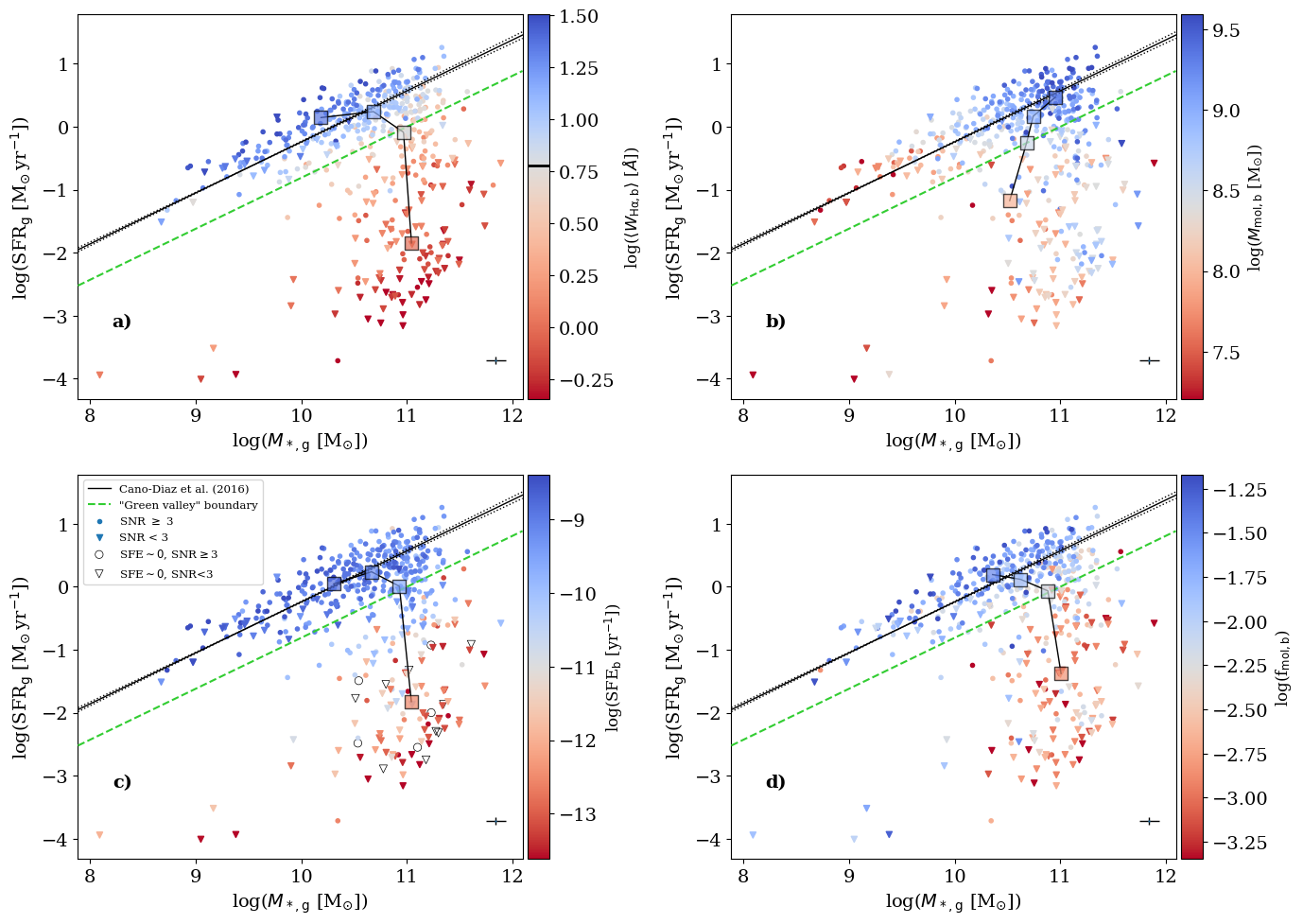}
    \caption{SFR-$M_*$ diagrams integrated over the CALIFA FoV, colour-coded by the median of the following quantities calculated in the APEX beam as described in the text: (a) median H$\alpha$ equivalent width  ($\langle W_{\rm H\alpha,b}\rangle$), (b) molecular gas mass ($M_{\rm mol, b}$), (c) star formation efficiency (SFE$_{\rm b}$) and (d) fraction of molecular gas with respect to the stellar mass ($f_{\rm mol, b}$). The solid black line indicates the star formation main sequence fit by Cano-D\'{i}az et al. (2016) with its confidence level (dotted lines). The green dashed line is $3\sigma$ (0.6 dex) below the SFMS fit, which we assume indicates the start of the ``green valley''. In each panel, circles indicate CO detections (SNR$\ge$3), and triangles non-detections (SNR$<$3). In panel $c$, unfilled symbols show data with SFE$\sim0$ (i.e. SFR$\sim0$) within the APEX beam aperture (see text for further details). The squares illustrate the position of the average $M_*$ and SFR at four different percentile ranges of the colouring parameter (<25\%, 25-50\%, 50-75\% and >75\%), with their colours indicating the average value at each percentile range.  The error-bar in the bottom-right of each panel shows the average errors of the reported parameters. The black horizontal line in panel $a$ colour-bar indicates the demarcation $\langle W_{\rm H\alpha,b}\rangle=6\,\AA$ value in logarithmic units.}
    \label{F:sfm}
\end{figure*}

\subsection{IFS-derived parameters}\label{SS:props_optical}

For the purpose of this work, we make use of both nebular lines as well as stellar continuum derived maps provided by CALIFA data. In particular we use \ha, \hb, [OIII] $\lambda5007$, [NII] $\lambda6583$ flux maps, $F_{\rm H\alpha}$, $F_{\rm H\beta}$, $F_{\rm [OIII]}$, $F_{\rm [NII]}$, respectively; the \ha\ equivalent width, \wha, and the stellar mass surface density maps.

We calculate the extinction-corrected star formation rate (SFR) spaxel-by-spaxel using the nebular extinction based on the Balmer decrement:

\begin{equation}\label{E:Aha}
    A_{\rm H\alpha}\,\mathrm{[Mag]} = \frac{K_{\rm H\alpha}}{0.4(K_{\rm H\beta} - K_{\rm H\alpha})}\times \log\left(\frac{F_{\rm H\alpha}}{2.86 F_{\rm H\beta}}\right),
\end{equation}

where the coefficients $K_{\rm H\alpha}=2.53$ and $K_{\rm H\beta}=3.61$ follow the \cite{cardelli1989} extinction curve \citep[see also ][]{catalan_torrecilla2015}. 

The SFR is then computed as:

\begin{equation}\label{E:sfr}
    \mathrm{SFR}\,\mathrm{[M_{\odot}\,yr^{-1}]} = 8\times 10^{-42} F_{\rm H\alpha}\times 10^{A_{\rm H\alpha}/2.5},
\end{equation}

as indicated by \cite{kennicutt1998} which assumes the Salpeter initial mass function \citep{salpeter1955}. To obtain the integrated SFR within the APEX beam aperture we coadded all  spaxels where \wha $> 6\AA$. In regions where  \wha $< 6\AA$ the \ha\ flux is not due to recent star formation but is dominated by the old-stellar population or other effects \citep{sanchez2013,espinosa-ponce2020}. Additionally, we remove from the summation all  spaxels where the ionisation is due to AGN, i.e. all the spaxels that fall above the BPT diagram \citep{baldwin1981} demarcation line given by \cite{kewley2001} in their equation 5. We distinguish between the beam SFR, SFR$_{\rm b}$=$\sum_i$ SFR$_i \times W_{{\rm T},i}$, as the integrated SFR within the APEX beam aperture, and the global SFR, SFR$_{\rm g}$=$\sum_i \mathrm{SFR}_i$, as the co-addition of  pixels over the whole map. 

The beam stellar mass, $M_{\rm *,b}$=$\sum_i$ $M_{{\rm *},i}\times W_{{\rm T},i}$, is obtained in the same way by the summation of the stellar masses from the spaxels within the APEX beam aperture, while the global stellar mass, $M_{\rm *,glob}$=$\sum_i$ $M_{{\rm *},i}$, is given by the summation over the whole map. In this formulae, the index $i$ runs over the spaxels of the whole maps, and the quantities without sub-script indicate the respective CALIFA data-derived maps.

To calculate $\alpha_{\rm CO}$ within the APEX beam aperture we measure the gas-phase metallicity over the CALIFA maps using the O3N2 method \citep{marino2013}.

\begin{equation}\label{E:metallicity}
    12 + \log\left(\frac{\rm O}{\rm H}\right) = 8.533 - 0.214\times \log\left(\frac{F_{\rm [OIII]}}{F_{\rm H\beta}} \frac{F_{\rm H\alpha}}{F_{\rm [NII]}}\right).
\end{equation}

As before the median gas-phase metallicity within the APEX beam aperture is calculated assuming the weights given by the Gaussian filter, $W_{\rm T}$. Using this method we obtain a median of 8.43 dex from the full galaxy sample with an inter-quartile range of 0.07 dex. The median metallicity with respect to the Solar metallicity \citep[8.69; ][]{allende_prieto2001} is $Z'_{\rm b}=0.55$.

Lastly, we calculate the star formation efficiency (SFE) and the molecular gas mass fraction (with respect to the stellar mass) within the APEX beam aperture, respectively, using:

\begin{equation}\label{E:SFE}
    \mathrm{SFE_b}\,[yr^{-1}] = \mathrm{SFR_b}/M_{\rm mol,b},
\end{equation}

\begin{equation}\label{E:fmol}
    f_{\rm mol,b} = M_{\rm mol,b}/M_{\rm *,b}.
\end{equation}

\begin{figure*}
    \centering
    \includegraphics[width = 0.85 \paperwidth, keepaspectratio]{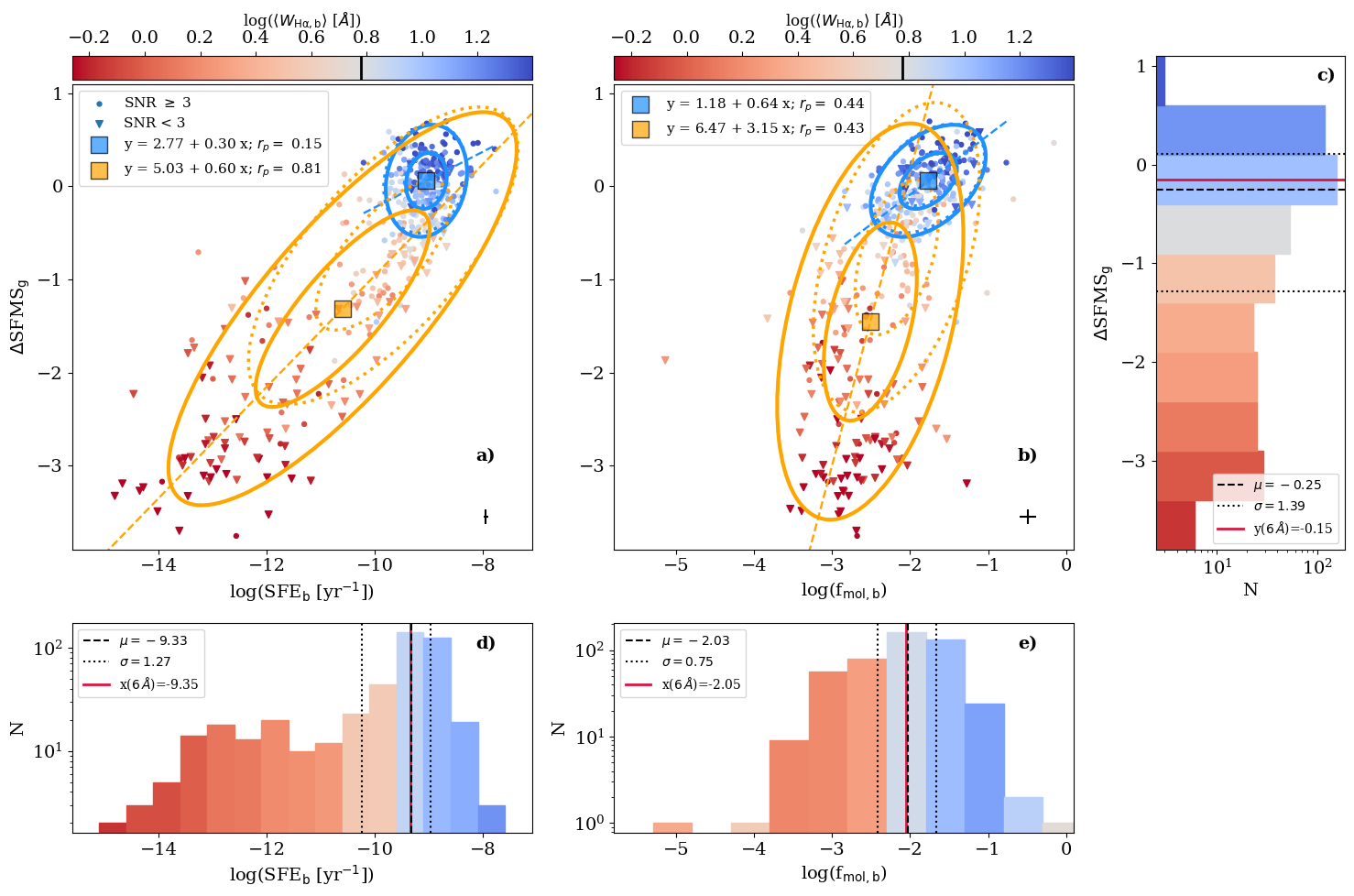}
    \caption{Offset from the main sequence for a galaxy ($\Delta$SFMS$_{\rm g}$) versus star formation efficiency  (SFE$_{\rm b}$, panel $a$) and versus the molecular gas fraction inside the APEX beam ($f_{\rm mol, b}$, panel $b$) colour-coded by the median H$\alpha$ equivalent width ($\langle W_{\rm H\alpha, b} \rangle$) within the APEX beam aperture. In the two panels, circles represent  CO detections (SNR$\ge3$), while triangles indicate CO upper limits (SNR$<3$). Error-bars at the bottom-right of each figure represent the typical uncertainties of the represented parameters. The two sub-samples include galaxies largely retired in the centre ($\langle W_{\rm H\alpha, b}\rangle <6\AA$), or dominated by star formation ($\langle W_{\rm H\alpha, b}\rangle >6\AA$), following the results of Fig~\ref{F:sfm}. The squares represent the median values of the represented parameters for the two sub-samples, while the ellipses correspond to the shape of the distribution derived using the PCA analysis described in the text, and contain approximately  1 and 2$\sigma$ of the data within the two sub-samples. Dotted-line ellipses are obtained including only CO detections, while solid-line ellipses correspond to the full sample (which also includes CO upper limits). Due to the different dynamical range of the x- and y-axes, some ellipses could result distorted, therefore the dashed coloured lines clarify the principal component direction for the full sub-samples. In the legend, the formulas indicate the linear fits derived from the two sub-samples using the PCA analysis, and $r_p$ the respective Pearson correlation coefficients. Panels $c$, $d$, and $e$ show the histogram distributions of $\Delta$SFMS$_{\rm g}$, SFE$_{\rm b}$, and $f_{\rm mol,b}$, respectively colour-encoded by the average $\langle W_{\rm H\alpha, b}\rangle$ in each bin. The dashed line indicates the median of the distribution ($\mu$), while the dotted lines the interquartile range ($\sigma$). The red line indicates the value of the quantity at the demarcation value given by $\langle W_{\rm H\alpha, b}\rangle =6\AA$ (i.e., the separation between centrally star-forming and retired galaxies). The black vertical line in  the panel $a$ and $b$ colour-bars indicates the demarcation $\langle W_{\rm H\alpha,b}\rangle=6\,\AA$ value in logarithmic units.}
    \label{F:dsfr}
\end{figure*}

\section{Results}\label{S:results}

\subsection{The SFR - $M_*$ diagram}\label{SS:sfm}

Fig.~\ref{F:sfm} displays the distribution of galaxies across the SFR-$M_*$ diagram using SFR and $M_*$ measurements integrated over the entire CALIFA field of view (FoV), SFR$_{\rm g}$ and $M_{\rm *, g}$, respectively. Each panel illustrates a different quantity calculated over the APEX beam aperture.  Quantities shown include: a) equivalent width of H$\alpha$ ($\langle W_{\rm H\alpha, b}\rangle$), b) molecular gas mass ($M_{\rm mol, b}$), c) star formation efficiency (SFE$_{\rm b}$), and d) the ratio of molecular gas mass to stellar mass (molecular gas mass fraction $f_{\rm mol, b}$), respectively. On average, the CALIFA maps extend to approximately 2\,$R_{\rm eff}$, while the APEX beam covers the inner 1\,$R_{\rm eff}$ of the galaxies (see Fig.~\ref{F:mapspectra}). The measurements within the APEX beam aperture can be considered, therefore, as inner galaxy measurements. In addition, we include the SFMS fit derived by \cite{cano_diaz2016} ($\log(SFR) = (0.81\pm0.02)\log(M_*) - (8.34\pm0.19)$), to illustrate the location of the star-forming galaxies across this diagram. In the following, we will provide the Pearson and Spearman correlation coefficients $r$ as $r_p$ and $r_s$, respectively. For all correlation discussed in the paragraphs we obtain extremely low $p-$value$<<10^{-16}$.

Global star-forming galaxies can be largely separated from galaxies on the way to quenching using a threshold given by $\langle W_{\rm H\alpha, b}\rangle=6\,\AA$, i.e. by considering the quenching stage of their centre. Panel $a$ of Fig.~\ref{F:sfm}, where the data points are colour-encoded by $\langle W_{\rm H\alpha, b}\rangle$, shows that most galaxies where this value is above 6$\,\AA$ are tightly distributed along the SFMS locus defined by the \cite{cano_diaz2016} fit. We consider the lower boundary of the SFMS to be $3\sigma$ ($\sim0.6\,$dex) below the fit, which encompasses $\sim99\%$ of the galaxies dominated by star-formation in their centre (which show similar medians $\langle W_{\rm H\alpha, b}\rangle$ and $\langle W_{\rm H\alpha, g}\rangle$ around $13\AA$). However, we still observe $\sim25\%$ of the centrally retired galaxy sub-sample (having $\langle W_{\rm H\alpha, b}\rangle<6\,\AA$) above this line. Those galaxies have median $\langle W_{\rm H\alpha, b}\rangle\sim4.3\,\AA$ and median $\langle W_{\rm H\alpha, g}\rangle\sim5.3\,\AA$, quite close to our threshold of 6$\,\AA$ and are mostly star-forming in their outskirts but not in their centres. 

It is interesting to note that quenched objects still possess a significant amount of molecular gas in their centres (Fig.~\ref{F:sfm}, panel $b$). In particular, we see objects with high or average values of this quantity well within the ``green valley'' and the ``red sequence''; for example 22 galaxies with $M_{\rm mol,b}$ values above the 75$^{\rm th}$ percentile of its distribution are below $3\sigma$ from the SFMS fit. At the opposite extreme, 46 objects with $M_{\rm mol,b}$ values below the 25$^{\rm th}$ percentile of the $M_{\rm mol,b}$ distribution are above the dashed green line defining galaxies within $3\sigma$ of  the SFMS fit. Fig.~\ref{F:sfm} (panel $b$) shows that $M_{\rm mol,b}$ is directly related to the global SFR and $M_*$, but correlates more tightly with SFR$_{\rm g}$ (given $r_p=0.65$ and $r_s=0.73$) than with $M_{\rm *, g}$ ($r_p=0.46$ and $r_s=0.33$).

The SFE in the centre of the galaxies varies widely across the SFR-$M_*$ diagram. However, it appears roughly constant along the SFMS, but it drops sharply right below it (panel $c$). Quantitatively, $\mu_{\rm SFE_b}=8.85\times10^{-10}\,$yr$^{-1}$ and $\sigma_{\rm SFE_b}=0.10\times10^{-10}\,$yr$^{-1}$ above the ``green valley'' boundary, while below it $\mu_{\rm SFE_b}=1.80\times10^{-10}\,$yr$^{-1}$ and $\sigma_{\rm SFE_b}=0.62\times10^{-10}\,$yr$^{-1}$. In other words, galaxies below the main sequence show, on average, depletion times a factor 5 longer than galaxies across the SFMS in their central regions. Additionally, we have a few galaxies that do not appear to form stars in their centre ($R<R_{\rm eff}$), resulting in SFE$\sim0$ in the aperture defined by the APEX beam. Those are galaxies with \ha\ flux below the detection limit, or their \ha\ map spaxels are masked since \wha\ $<6\,\AA$. Nevertheless, most of these galaxies do have a few star-forming regions in the outskirts, typically outside the circle defined by their R$_{\rm eff}$ (such as NGC0171 in Fig.~\ref{F:mapspectra}). Some of these galaxies are non-detected in CO (9 targets); therefore the absence of recent star formation could be directly attributed to the absence of molecular gas (in the matched aperture). However, some of the SFE$\sim0$ galaxies are detected in CO (5 targets), meaning that the star formation quenching is the centre is due to causes other than a simple shortage of raw fuel. As for $M_{\rm mol,b}$, we also observe a few galaxies with SFE$_{\rm b}$ much lower than the sample average very close to the main sequence (5 targets with SFE$_{\rm b}$ below the 25$^{\rm th}$ percentile of the SFE distribution are above the $3\sigma$ line). SFE$_{\rm b}$ decreases with stellar mass (as observed also in \citealt{colombo2018} with a limited sample of EDGE galaxies; see their Fig. 3, panel $d$). The SFE in the galaxy centres appear quite strongly correlated with SFR$_{\rm g}$ (given $r_p=0.73$ and $r_s=0.46$), but only moderately with $M_{\rm *, g}$ ($r_p=-0.38$ and $r_s=-0.53$). The SFE$_{\rm b}$ reaches very low values, however, those values are generally SFE$_{\rm b}$ lower limits, being mostly driven by CO non-detected galaxies, for which we use an $M_{\rm mol, b}$ upper limit to calculate SFE$_{\rm b}$. 

The ratio of the molecular gas mass to stellar mass $f_{\rm mol,b}$ shows a general behaviour across the SFR-$M_*$ diagram quite similar to SFE$_{\rm b}$ (panel $d$). As for  SFE$_{\rm b}$, $f_{\rm mol,b}$ is largely constant along the SFMS and sharply decreases below from the SFMS. Quantitatively, $\mu_{ f_{\rm mol,b}}=2\times10^{-2}$ and $\sigma_{f_{\rm mol,b}}=2\times10^{-2}$ above the ``green valley'' boundary, and $\mu_{ f_{\rm mol,b}}=5\times10^{-3}$ and $\sigma_{f_{\rm mol,b}}=9\times10^{-3}$ below it. We observe galaxies along and below the main sequence with $f_{\rm mol,b}$ values much lower and much higher than the sample average. In particular we have 14 targets with $f_{\rm mol,b}$ below the 25$^{\rm th}$ percentile of the distribution above the $3\sigma$ line (dashed green line) from the SFMS fit, while 5 objects with $f_{\rm mol,b}$ values above the 75$^{\rm th}$ are located below this line. Nevertheless, the ratio of molecular gas mass to stellar mass appears only moderately correlated with SFR$_{\rm g}$ ($r_p=0.50$ and $r_s=0.50$) and $M_{\rm *, g}$ ($r_p=-0.44$ and $r_s=-0.50$).

\begin{table*}
\setlength{\tabcolsep}{3.2pt}
\renewcommand{\arraystretch}{1.5}
\begin{tabular}{c|cccccc|cccccc}
\hline
& \multicolumn{12}{|c}{$\Delta$SFMS$_{\rm g}=q+m\log($SFE$_{\rm b})$} \\
\hline
Type & $m$ & $q$ & $\langle\log($SFE$_{\rm b})\rangle$ & $\langle \Delta$SFMS$_{\rm g}\rangle$ & $r_p$ & $r_s$ & $m$ & $q$ & $\langle\log($SFE$_{\rm b})\rangle$ & $\langle \Delta$SFMS$_{\rm g}\rangle$ & $r_p$ & $r_s$ \\
\hline
& \multicolumn{6}{|c}{Centrally star-forming, $\langle W_{\rm H\alpha,b}\rangle>6\,\AA$} & \multicolumn{6}{|c}{Centrally retired, $\langle W_{\rm H\alpha,b}\rangle<6\,\AA$} \\
\hline
Best & $0.30^{+0.03}_{-0.03}$ & $2.77^{+0.30}_{-0.27}$ & -9.05 & 0.06 & 0.15 & 0.05 & $0.60^{+0.01}_{-0.01}$ & $5.03^{+0.10}_{-0.09}$ & -10.60 & -1.32 & 0.81 & 0.81 \\
SNR>3 & $0.27^{+0.03}_{-0.03}$ & $2.56^{+0.23}_{-0.25}$ & -9.07 & 0.08 & 0.16 & 0.05 & $0.54^{+0.02}_{-0.01}$ & $4.61^{+0.16}_{-0.13}$ & -9.85 & -0.75 & 0.71 & 0.63 \\
Const. $\alpha_{\rm CO}$ & $0.26^{+0.02}_{-0.02}$ & $2.42^{+0.20}_{-0.20}$ & -9.19 & 0.06 & 0.19 & 0.13 & $0.60^{+0.03}_{-0.01}$ & $5.12^{+0.30}_{-0.11}$ & -10.50 & -1.14 & 0.80 & 0.79 \\
Only APEX & $0.27^{+0.03}_{-0.03}$ & $2.52^{+0.27}_{-0.27}$ & -9.00 & 0.08 & 0.16 & 0.05 & $0.57^{+0.01}_{-0.01}$ & $4.56^{+0.11}_{-0.10}$ & -10.86 & -1.59 & 0.83 & 0.84 \\
\hline
& \multicolumn{6}{|c}{Globally star-forming, $\langle W_{\rm H\alpha,g}\rangle>6\,\AA$} & \multicolumn{6}{|c}{Globally retired, $\langle W_{\rm H\alpha,g}\rangle<6\,\AA$} \\
\hline
Best & $0.71^{+0.03}_{-0.03}$ & $6.47^{+0.31}_{-0.28}$ & -9.10 & 0.02 & 0.46 & 0.24 & $0.60^{+0.01}_{-0.01}$ & $5.19^{+0.10}_{-0.11}$ & -11.19 & -1.51 & 0.80 & 0.81 \\
\hline
\hline
& \multicolumn{12}{|c}{$\Delta$SFMS$_{\rm g}=q+m\log(f_{\rm mol,b})$} \\
\hline
Type & $m$ & $q$ & $\langle\log(f_{\rm mol,b})\rangle$ & $\langle \Delta$SFMS$_{\rm g}\rangle$ & $r_p$ & $r_s$ & $m$ & $q$ & $\langle\log(f_{\rm mol,b})\rangle$ & $\langle \Delta$SFMS$_{\rm g}\rangle$ & $r_p$ & $r_s$ \\
\hline
& \multicolumn{6}{|c}{Centrally star-forming, $\langle W_{\rm H\alpha,b}\rangle>6\,\AA$} & \multicolumn{6}{|c}{Centrally retired, $\langle W_{\rm H\alpha,b}\rangle<6\,\AA$} \\
\hline
Best & $0.64^{+0.03}_{-0.02}$ & $1.18^{+0.05}_{-0.04}$ & -1.77 & 0.06 & 0.44 & 0.51 & $3.15^{+0.07}_{-0.06}$ & $6.47^{+0.19}_{-0.15}$ & -2.51 & -1.45 & 0.43 & 0.48 \\
SNR>3 & $0.63^{+0.03}_{-0.03}$ & $1.19^{+0.05}_{-0.05}$ & -1.76 & 0.08 & 0.42 & 0.51 & $2.40^{+0.08}_{-0.07}$ & $4.45^{+0.18}_{-0.16}$ & -2.17 & -0.76 & 0.46 & 0.53 \\
Const. $\alpha_{\rm CO}$ & $0.86^{+0.04}_{-0.04}$ & $1.46^{+0.06}_{-0.06}$ & -1.63 & 0.06 & 0.43 & 0.51 & $3.32^{+0.06}_{-0.06}$ & $6.06^{+0.15}_{-0.14}$ & -2.18 & -1.18 & 0.52 & 0.59 \\
Only APEX & $0.59^{+0.03}_{-0.03}$ & $1.16^{+0.05}_{-0.05}$ & -1.82 & 0.08 & 0.45 & 0.53 & $4.14^{+0.15}_{-0.13}$ & $9.12^{+0.40}_{-0.34}$ & -2.66 & -1.86 & 0.32 & 0.34 \\
\hline
& \multicolumn{6}{|c}{Globally star-forming, $\langle W_{\rm H\alpha,g}\rangle>6\,\AA$} & \multicolumn{6}{|c}{Globally retired, $\langle W_{\rm H\alpha,g}\rangle<6\,\AA$} \\
\hline
Best & $1.31^{+0.04}_{-0.04}$ & $2.40^{+0.08}_{-0.08}$ & -1.82 & 0.02 & 0.45 & 0.53 & $3.55^{+0.10}_{-0.08}$ & $7.49^{+0.26}_{-0.21}$ & -2.58 & -1.67 & 0.39 & 0.40 \\
\hline
\end{tabular}
\caption{Summary of the PCA fit of the $\Delta$SFMS$_{\rm g}$-$\log($SFE$_{\rm b})$ and $\Delta$SFMS$_{\rm g}$-$\log(f_{\rm mol,b})$ relationships for centrally star-forming ($\langle W_{\rm H\alpha,b}\rangle>6\,\AA$) and centrally retired ($\langle W_{\rm H\alpha,b}\rangle<6\,\AA$) (or globally star-forming, $\langle W_{\rm H\alpha,g}\rangle>6\,\AA$, and globally retired, $\langle W_{\rm H\alpha,g}\rangle<6\,\AA$) for different galaxy sub-samples or quantity calculations. ``Best'' indicates the whole galaxy datasets with molecular gas masses calculated using a variable $\alpha_{\rm CO}$, which results are shown in Fig.~4; for ``SNR>3'' only the detections are considered; ``Const. $\alpha_{CO}$, represents the whole galaxy sample, where a constant $\alpha_{\rm CO}$ is used to convert CO luminosities into molecular gas masses; ``Only APEX'' marks the fit results when only the APEX data are used. In the columns, $m$ and $q$ indicate slope and intercept of the relations, respectively, inferred from PCA; $\langle \Delta$SFMS$_{\rm g}\rangle$, $\langle\log($SFE$_{\rm b})\rangle$, and $\langle\log(f_{\rm mol,b})\rangle$, show the medians of global $\Delta$SFMS , beam SFE, and beam \fmol\ of the distributions, respectively; $r_p$ and $r_s$ are the Pearson and Spearman correlation coefficients, respectively. For most of the correlation realizations we obtain $p-$values largely below $10^{-5}$ from both correlation tests, except for $\Delta$SFMS$_{\rm g}$-SFE$_{\rm b}$ relations for the centrally star-forming galaxies, for which we measure Pearson $p-$values of the order of $10^{-2}$ and Spearman $p-$values of the order of $10^{-1}$. The uncertainties are obtained by 1000 bootstrap iterations of the PCA fit, and are provided as 75$^{th}$-50$^{th}$ percentiles and 50$^{th}$-25$^{th}$ percentiles of the $m$ and $q$ distributions.}
\label{T:pca_fit}
\end{table*}

\subsection{What quenches galaxies: variable SFE or shortage of molecular gas?
}\label{SS:dsfr}

Star formation quenching can be parameterised using the logarithmic difference between the observed SFR and the SFR expected from the best fit to the SFMS, $\Delta$SFMS \citep[e.g., ][]{genzel2015,tacconi2018,ellison2018,thorp2019}. In panels $a$ and $b$ of Fig.~\ref{F:dsfr} we plot \dms\ with respect to  SFE$_{\rm b}$ and $f_{\rm mol,b}$, respectively, in order to understand whether the star formation quenching is more tightly connected to variations in SFE or to the absence of molecular gas in galaxy centres. This is basically a reorganisation of the star formation-mass diagram presented in Fig.~\ref{F:sfm}, removing the dependence of SFR on $M_*$ (i.e., the SFMS trend). As before, SFE$_{\rm b}$ and $f_{\rm mol,b}$ are measured within the APEX beam aperture, while \dms\ uses the SFR and $M_*$ measured over the entire CALIFA map. Following the arguments discussed for Fig.~\ref{F:sfm} (panel $a$), we divide the sample into two sub-samples based on the average \wha\ within the APEX beam aperture ($\langle W_{\rm H\alpha, b}\rangle$): galaxies largely quenched in the centre ($\langle W_{\rm H\alpha, b} \rangle<6\AA$) and galaxies dominated by star formation in their centres( $\langle W_{\rm H\alpha, b}\rangle >6\AA$). The two sub-samples are well balanced in terms of target size. Centrally star-forming galaxies number 256, i.e. $\sim54\%$ of the full sample, while the centrally retired galaxies constitute $\sim46\%$ of the sample, i.e. 216 objects.

The behaviour of  \dms\ versus SFE$_{\rm b}$ is somehow similar for the two sub-samples. The \dms - SFE$_{\rm b}$ relationship measured using the principal component analysis (PCA, see \citealt{colombo2018}) shows that the slope from the confidence ellipsoids between the two sub-samples is on the same order $\sim0.3$ for centrally star-forming galaxies and $\sim0.6$ for centrally quenched galaxies (see Table~\ref{T:pca_fit}, where also Spearman correlation coefficient $r_s$ for the two sub-samples are reported). 

Nevertheless, data points for star-forming galaxies are tightly concentrated close to the SFMS and have SFE$_{\rm b}$ values between $10^{-10}-10^{-8}$\,yr$^{-1}$ (i.e. $\tau_{\rm dep}=0.1-10\,$Gyr). By contrast, quenched galaxies cover a much larger parameter space in both \dms\ and SFE$_{\rm b}$, in particular, they span 6 orders of magnitudes in SFE. Additionally, \dms\ and SFE$_{\rm b}$ appear strongly correlated, showing a Pearson $r_p=0.9$. However, this tight correlation is mostly driven by the centrally quenched galaxies, for which $r_p=0.8$, while for the star-forming targets $r_p=0.2$, which indicates that \dms\ and SFE$_{\rm b}$ are basically uncorrelated for this kind of object and the calculated slope of the relationship is meaningless. SFE in our centrally star-forming galaxies is quite constant, in line with results from several other resolved and unresolved studies of nearby, star-forming galaxies. Note also that the values of the correlation coefficients do not change significantly for the \dms\ - SFE$_{\rm b}$ and \dms\ - $f_{\rm mol,b}$ relationships if only the detected targets are considered (Table~\ref{T:pca_fit}).

On the other hand, the slopes of relationship between \dms\ and $f_{\rm mol,b}$ are starkly different if we consider the star-forming and the quenched targets separately. Galaxies largely quenched in the centre span a few orders of magnitude in \dms\ as well as in $f_{\rm mol,b}$. Indeed, the PCA shows that in quenched galaxies there is a steep relationship between \dms\ and $f_{\rm mol,b}$ (with a slope of $\sim3.15$). However, this correlation is shallower for galaxies with central star-formation activity (slope $\sim0.64$). They have $f_{\rm mol,b}$ approximately one order of magnitude larger than for centrally quenched galaxies ($\langle \log(f_{\rm mol,b}) \rangle = -1.77$ for centrally star-forming and $\langle \log(f_{\rm mol,b}) \rangle = -2.51$ for centrally retired objects; see Table~\ref{T:pca_fit}). 

Nevertheless, the Pearson correlation coefficients are lower with respect to the SFE case: for the full sample, we observe $r_p=0.7$, while for the two sub-samples separately we observe a similar $r_p\sim0.5$. This indicates that, in contrast to the SFE case, for both centrally star-forming and quenched targets \dms\ and $f_{\rm mol,b}$ are moderately correlated. Those conclusions do not change significantly if only the CO detected galaxies are considered (see Fig.~\ref{F:dsfr} and Table~\ref{T:pca_fit}).

It is worth noting that the SFE$_{\rm b}$ exhibits a bimodal distribution similar to the one found for \dms\ , i.e. values of SFE$_{\rm b}$ below $10^{-10}\,$yr$^{-1}$ are almost exclusively associated with quenched targets. By contrast, a bimodal distribution is not evident for $f_{\rm mol,b}$, for which the difference in $f_{\rm mol,b}$ between centrally star-forming and retired objects is not as sharp. Additionally, the bimodality in SFE$_{\rm b}$ and \dms\ is driven by the same group of galaxies. In other words, centrally star-forming and retired galaxy sub-groups are equally well separated in \dms\ and in SFE$_{\rm b}$. Thus, SFE in the galaxy centres (in particular retired centres) is a better predictor of the separation between the two groups than the respective $f_{\rm mol,b}$.

Panels $c$, $d$, and $e$ show the histograms of \dms, SFE$_{\rm b}$, and $f_{\rm mol,b}$ colour-coded by \mwhab\ in a given bin. Generally, the median of global \dms\ ($-0.3$) and beam SFE$_{\rm b}$ ($\sim4.4\times10^{-9}$\,yr) are close to the values of these parameters that separate star-forming and quenched galaxies (i.e., $\Delta$SFMS($W_{\rm H\alpha}=6\AA$) and SFE($W_{\rm H\alpha}=6\AA$)). In particular, the median SFE corresponds to a $\tau_{\rm dep}=2.3\,$Gyr, which is equivalent to the value measured from kpc-resolved EDGE objects \citep[see ][]{utomo2017,colombo2018} and other nearby spiral galaxies \citep{leroy2013}. However, the median of the beam $f_{\rm mol,b}$ ($\sim10^{-2}$) distribution is shifted towards the retired sub-sample as this value is slightly below the demarcation $f_{\rm mol,b}$ that separates centrally star-forming and quiescent galaxies ($f_{\rm mol,b}$($W_{\rm H\alpha}=6\AA$)=$10^{-1.95}$).

\section{Discussion and conclusions}\label{S:summary}
In this paper, we use 472 galaxies to test whether the star formation quenching of CALIFA galaxies is mostly due to changes in the SFE or to the absence of molecular gas (as described by the ratio between the molecular and stellar gas masses, \fmol) in their centres. 

\emph{We observe that for galaxies dominated by star formation activity in their centre, distance from the main sequence correlates better with the molecular to stellar mass ratio. For centrally quiescent galaxies, instead, distance from the main sequence correlates better with SFE. This suggests a scenario where the progressive loss of the cold gas reservoir is what causes galaxies to move out of the main sequence. Once this happens, the star formation efficiency in the remaining cold gas reservoir is what modulates their retirement, with lower efficiencies corresponding to more quiescent galaxies. In this scenario both amount of (molecular) gas and SFE matter, but they have different roles. In particular, the stabilisation of the molecular gas reservoir plays a role once the galaxy enters the green valley, but it is less important than the size of the reservoir to move the galaxy out of the main sequence. Furthermore, this quenching happens from the inside-out, with centrally quenched galaxies leading the path towards totally quenched ones.}

Those results do not change significantly if we consider a constant $\alpha_{\rm CO}$ instead of our preferred $\alpha_{\rm CO}$ from Eq.~\ref{E:aco} in converting CO luminosity to molecular gas mass, or if we divide the full sample using the value of the H$\alpha$ equivalent width obtained over the full maps, or if only the APEX sub-sample of galaxies is used (see Table~\ref{T:pca_fit}). 

The importance of the absence of molecular gas for understanding why some galaxies are located far from the SFMS has been acknowledged in the past from other (integrated, but aperture limited) studies that used a direct molecular gas tracer as CO, in both local and higher redshift Universe. At $z\sim0$, a series of papers using the COLDGASS\footnote{``CO Legacy Database
for the GASS'' survey} and xCOLDGASS\footnote{``Extended CO Legacy Database
for the GASS'' survey} samples \citep{saintonge2012,saintonge2016,saintonge2017} have shown that variations of the specific star formation rate (sSFR=SFR/$M_*$) can be almost fully described by variations in gas fractions (especially molecular gas fraction), but the relation between \fmol\ and sSFR is not linear, meaning that variations in star formation efficiency (which appears almost constant with the stellar mass, cf. \citealt{saintonge2016}) also plays a role. Similar results are obtained by extending the sample to higher redshift \citep[up to $z\sim4$ ][]{genzel2015}. 

A relatively inexpensive way to explore the distribution of molecular gas in galaxies is to use indirect proxies. In particular, the dust-to-gas relation can be applied to estimate both the integrated $M_{\rm mol}$ and its distribution across galaxies. A recent calibrator proposed by \cite{barrera-ballesteros2020} was used in \cite{sanchez2018} and \cite{lacerda2020} to explore the radial distribution of the molecular gas and its integrated molecular gas mass for different galaxy morphologies. They confirm the results by \cite{colombo2018}, in terms of the variation of the SFE across galaxy types and stellar masses, despite the limitations of the adopted estimators. Furthermore, \cite{sanchez2018} using two large samples of IFS spatially-resolved observations comprising ~2700 galaxies from the MaNGA\footnote{``Mapping Nearby Galaxies at Apache Point Observatory''} \citep{bundy2015} IFS survey (and ~8000 galaxies from a large IFS compilation) confirm that the SFE decreases as galaxies move from the MS to the retired galaxies regime, going through the green valley (see their Fig. 8 and 11), as recently reviewed by \cite{sanchez2020} (their Fig. 18). Like in the case of the (x)COLDGASS results, they attribute to the lack of gas and not the low SFE, the primary cause of the cessation of star formation.

Similarly, \cite{piotrowska2020}, using a dust-to-gas calibrator method to analyse $\sim62,000$ SDSS DR7 local galaxies, find that (independently from the stellar mass) both decreasing gas supply and decreasing efficiency are important to define the distance from the star formation main sequence, in line with the previously discussed integrated study results. \cite{tacconi2018} use both CO and dust-extrapolated molecular gas masses in the redshift range $z=0-4$ and also confirm the primary dependency of sSFR to \fmol\ with a weaker contribution from SFE changes \citep[see also ][]{scoville2016}. 

The same question has been recently addressed using spatially-resolved measurements. \cite{ellison2020} used 34 galaxies from the ALMaQUEST\footnote{``ALMA-MaNGA QUEnching and STar formation''} sample that images MaNGA targets in $^{12}$CO(1-0) with the Atacama Large Millimeter/submillimeter Array (ALMA) . This sample also includes green valley targets. They find that on kpc-scales, variations in SFE (measured as $\Sigma_{\rm SFR}/\Sigma_{\rm mol}$), rather than resolved \fmol\ changes (calculated from $\Sigma_{\rm mol}/\Sigma_*$), drive the SFR surface density of galaxies away from the ``resolved'' star formation main sequence \citep{lin2019}. By analysing 7 ``green valley'' galaxies, \citep{brownson2020} found that SFE and \fmol\ appear equally important to explain quenching in the outer regions of galaxies. However, they were unable to establish which is the dominant mechanism in the galaxy centres, which appear strongly quenched in their sample. They indicated that, while low \fmol\ values seem to drive the quenching in the inner regions, reduced SFE could also play a role. Through a smaller sample of nearby galaxies, but observed at higher resolution, \cite{morselli2020} notice that changes in the total gas fraction (calculated including the contribution of the atomic gas) are more significant than the total SFE in explaining distance from the resolved SFMS for star-forming galaxies, as we observe here using integrated measurements.

Integrated CO surveys cannot reach the level of detail regarding the molecular gas organisation achieved by kpc-resolved studies. But they do provide the ability to obtain samples that are several times larger and to detect galaxies with much less molecular gas using much less observing time. In this paper, we give the first presentation of a new integrated CO survey using APEX that follows up CALIFA targets. Once completed, this survey will give $^{12}$CO(2-1) (and possibly also $^{13}$CO(2-1) and C$^{18}$O(2-1) for the brightest targets) observations of 450 CALIFA galaxy centres and a few off-centre detections. Thus, the size of this survey is similar to that of the most recent explorations at redshift $\sim0$, like xCOLDGASS (532 galaxies), but with aperture-matched optical spectroscopic data (not restricted to the central 3", which could cause several issues in the classification of the ionisation stages, \citealt{sanchez2020}), and for a much narrower range of cosmological distances (i.e., with less bias introduced by possible cosmological evolution). The survey is unbiased by construction, having as its only requirement that the targets are observable by APEX ($\delta<30^{\circ}$). Together with CARMA data, we will collect CO data for $\sim630$ galaxies fully covered by high-resolution IFS information, providing the largest CO database of any major IFS survey to date. 

Nonetheless, to fully exploit the IFS information and take the next step in understanding the mechanisms that drive these galaxy changes requires high-resolution interferometric gas imaging of the sample, something that needs to be strongly supported by proper time allocation. 

ALMA would be particularly appropriate for this scope. Within our centrally quenched sample, we measure a median molecular gas mass upper limit $M_{\rm mol}\sim10^8\,$M$_{\odot}$, within a 26.3 arcsec APEX beam. A short ($\sim20$~min) integration on CO(2-1) emission over the full disk of a CALIFA galaxy with ALMA 12m array would provide a $1\sigma$ sensitivity of $\sim8.7$\,mJy in 1\,km\,s$^{-1}$ channel. This is equivalent to $\Sigma_{\rm mol}\sim2$\,M$_{\odot}$\,pc$^{-2}$ for 30\,km\,s$^{-1}$-wide lines, which would correspond to a $M_{\rm mol}\simeq10^{6.2}\,$M$_{\odot}$ for the median distance to CALIFA galaxies, in a beam that matches the CALIFA resolution ($\sim3$ arcsec). Therefore a short integration with ALMA has a $5\sigma$ limit of $M_{\rm mol}\sim10^7$\,M$_\odot$ (depending on the precise distance and line-width of the target), and would be able to improve significantly on our limits and potentially resolve the faint CO emission of a quenched galaxy. At these integration times, large surveys are possible, so about $300$ CALIFA targets could be done in approximately 100 hours. This would provide a representative, invaluable high-resolution sample of galaxies to study the star formation quenching process in the local Universe.

\begin{acknowledgements}
The authors thank the anonymous referee for the constructive report. DC and AW acknowledges support by the \emph{Deut\-sche For\-schungs\-ge\-mein\-schaft, DFG\/} project number SFB956A. SFS thanks the support of Coancyt grants FC2016-01-1916 and CB-285080, and UNAM-DGAPA-PAPIIT IA100519. ER acknowledges the support of the Natural Sciences and Engineering Research Council of Canada (NSERC), funding reference number RGPIN-2017-03987. ADB, TW, LB, SV, and RCL acknowledge support from the National Science Foundation (NSF) through collaborative research award AST-1615960. TW and YC acknowledge support from the NSF through grant AST-1616199. JBB acknowledges support from the grant IA-100420 (PAPIIT-DGAPA, UNAM). This research made use of Astropy,\footnote{http://www.astropy.org} a community-developed core Python package for Astronomy \citep{astropy2013, astropy2018}; matplotlib \citep{matplotlib2007}; numpy and scipy \citep{scipy2020}. Support for CARMA construction was derived from the Gordon and Betty Moore Foundation, the Eileen and Kenneth Norris Foundation, the Caltech Associates, the states of California, Illinois, and Maryland, and the NSF. Funding for CARMA development and operations were supported by NSF and the CARMA partner universities.
\end{acknowledgements}

%
   \bibliographystyle{aa} 
   \bibliography{cold.bib} 
%

\end{document}